\documentclass[a4paper]{jpconf}
\usepackage{graphicx}
\usepackage{subfigure}
\begin{document}
\title{Heavy-flavor azimuthal correlations of $D$ mesons}

\author{Marlene Nahrgang$^{1,2,3}$, J\"org Aichelin$^{3}$, Pol Bernard Gossiaux$^{3}$, Klaus Werner$^{3}$}

\address{$^1$ Department of Physics, Duke University, Durham, North Carolina 27708-0305, USA}
\address{$^2$ Frankfurt Institute for Advanced Studies (FIAS), Ruth-Moufang-Str.~1, 60438 Frankfurt am Main, Germany}
\address{$^3$ SUBATECH, UMR 6457, Universit\'e de Nantes, Ecole des Mines de Nantes,
IN2P3/CNRS. 4 rue Alfred Kastler, 44307 Nantes cedex 3, France}

\ead{marlene.nahrgang@phy.duke.edu}

\begin{abstract}
Observables of heavy-quark azimuthal correlations in heavy-ion collisions are new and promising probes for the investigation of the in-medium energy loss. We explore the potential of these observables to discriminate the collisional and radiative contributions within a hybrid EPOS+MC@sHQ transport approach.
\end{abstract}

\section{Introduction}
Traditional heavy-quark observables like the nuclear modification factor $R_{\rm AA}$ and the elliptic flow $v_2$ are thoroughly measured at RHIC \cite{rhic} and the LHC \cite{Alice,delValle:2012qw,Abelev:2013lca}. The results suggest a significant energy loss at high transverse momentum $p_T$ and partial thermalization with the QGP medium at low $p_T$. Many theoretical approaches including energy loss via gluon bremsstrahlung 
\cite{radiative},
 elastic scatterings  \cite{elastic} or a mixture of both processes are able to reproduce the $R_{\rm AA}$ and $v_2$ data within numerical simulations, for which a rescaling of the interaction cross sections or of the diffusion coefficient and a coupling to a background fluid dynamical medium of light partons is necessary \cite{results,Gossiaux:2008jv,Gossiaux:2010yx,Gossiaux:2012ya}. It remains, however, a challenge to reproduce both of these observables within a single setup.
Besides the energy loss mechanism (elastic, radiative or a mixture of both) also the modeling of the fluid dynamical medium and the details of the coupling between the medium and the heavy quarks via an interpretation of the equation of state affect the numerical results substantially \cite{Gossiaux:2011ea,Nahrgang:2013xaa}. It is therefore of crucial importance to use fluid dynamical models, which are well tested in the light sector, and to properly couple the both sectors. 


We show that investigating heavy-flavor azimuthal correlations in addition to the $R_{\rm AA}$ and the $v_2$ offers the possibility to discriminate the different contributions to the energy loss, radiative or collisional. We use our approach of heavy-quark propagation, MC@sHQ coupled to a $3+1$~d ideal fluid dynamical evolution, which is subsequent to EPOS initial conditions. EPOS in its integral version, i.~e. including a hadronic afterburner, successfully reproduces light-hadron observables \cite{EPOS}. The heavy quarks are initialized via the $p_T$-distributions from FONLL \cite{fonll}, then propagated via the Boltzmann equation taking the local temperature and fluid velocities from the background medium and finally hadronized via coalescence at low-$p_T$ and fragmentation, which dominates at high-$p_T$ \cite{Gossiaux:2008jv,Gossiaux:2010yx,Gossiaux:2012ya,Nahrgang:2013saa,Nahrgang:2013xaa}.

Our procedure is the following: We will fix a global and temperature-independent $K$-factor by reproducing the LHC data of the high-$p_T$ $D$-meson $R_{\rm AA}$ within the experimental errors for each interaction mechanism. We will see that the collisional and radiative energy loss contribute about the same to the overall energy loss. With these respective $K$-factors we will then evaluate the $D$-meson $v_2$ and the azimuthal correlations of $D\bar{D}$-pairs.

\section{Properties of the interaction}

\begin{figure}[tb]
 \centering

   \subfigure{\label{fig:Adrag}\includegraphics[width=0.48\textwidth]{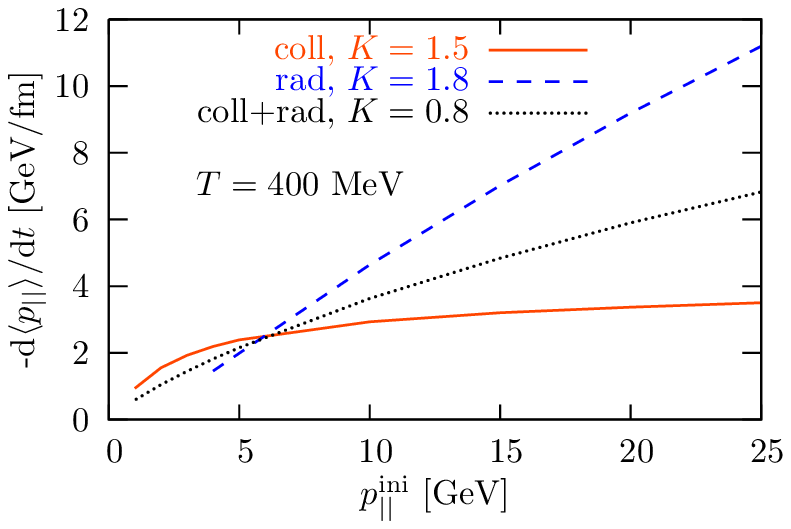}} \hfill
   \subfigure{\label{fig:pperp}\includegraphics[width=0.48\textwidth]{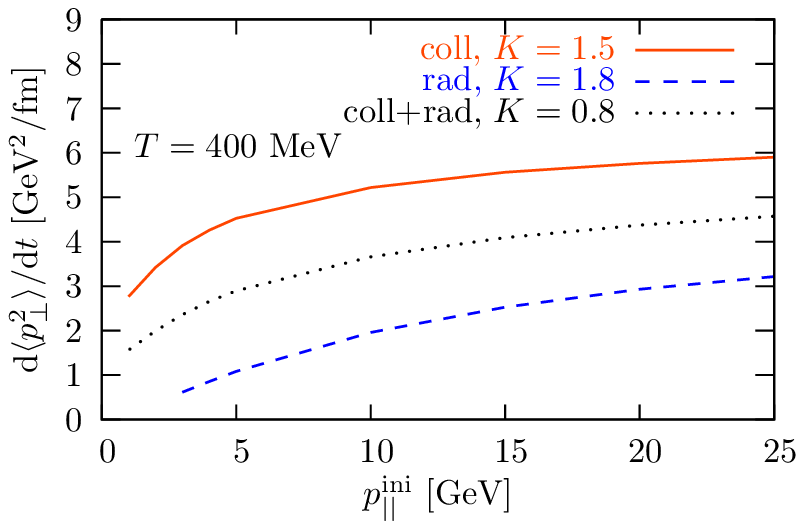}}
 \caption{The drag coefficient and the average perpendicular deflection for charm quarks.}
 \label{fig:prop}
\end{figure}

In a first step we analyze the drag coefficient and the average deflection of the charm quark with respect to its initial direction in an infinite, static medium at the given temperature $T=400$~MeV. For this purpose we initialize the charm quarks with $\vec{p}^{\,\rm ini}=(0,0,p_{||}^{\rm ini})$. The drag coefficient, which describes the average parallel momentum loss, is then given by $A=-{\rm d}\langle p_{||}\rangle/{\rm d}t$. We observe in figure \ref{fig:Adrag} that for the purely radiative energy loss the increase of $A$ with the charm quark momentum is much faster than for the purely collisional energy loss. The combination of collisional and radiative corrections lies between the two pure mechanisms. There is, however, a region for small and intermediate momenta, where the drag coefficient $A$ for the purely collisional energy loss mechanism is larger than for the purely radiative one. 

For the study of azimuthal correlations it is important to understand how much momentum perpendicular to its initial direction $p_\perp^2=p_x^2+p_y^2$ a charm quark acquires on average during the evolution. This quantity ${\rm d}\langle p_\perp^2\rangle/{\rm d}t$ is shown in figure \ref{fig:pperp}. Over the whole momentum range it is distinctively different for the three interaction mechanisms. It is largest for the purely collisional energy loss and smallest for the radiative energy loss. From this we expect that the initial azimuthal correlations will be broadened more efficiently by the purely collisional than by the radiative scatterings.

Due to approximations in the implementation the purely radiative scenario is not applicable for small transverse momenta $p_T<3$~GeV.

\section{Traditional observables}

\begin{figure}[tb]
 \centering

   \subfigure{\label{fig:RAA}\includegraphics[width=0.48\textwidth]{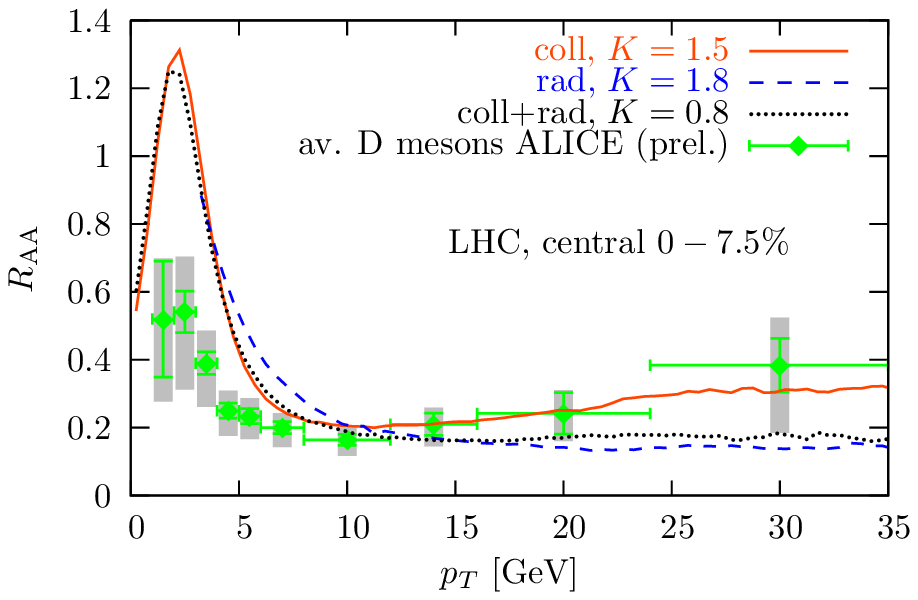}} \hfill
   \subfigure{\label{fig:v2}\includegraphics[width=0.47\textwidth]{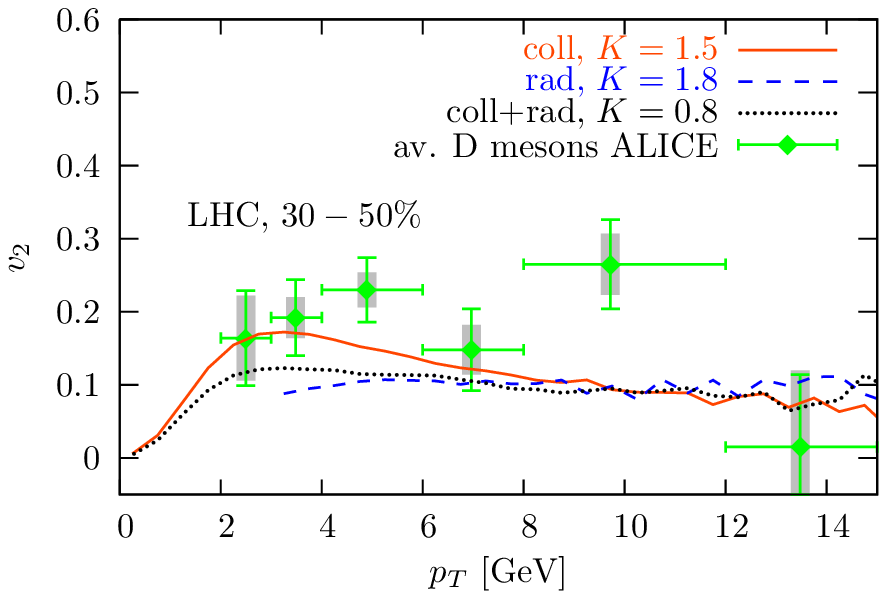}}
 \caption{The nuclear modification factor $R_{\rm AA}$ and the elliptic flow $v_2$ of average $D$-mesons.  ALICE data of averaged $D$ mesons, which includes $D^0$, $D^+$ and $D^{*+}$, is from \cite{delValle:2012qw} and \cite{Abelev:2013lca}.}
 \label{fig:RAAv2}
\end{figure}

For a decoupling temperature $T_c=155$~MeV we evaluate the D-meson $R_{\rm AA}$ in central $0-7.5$~\% and $v_2$ in mid-peripheral $30-50$~\% Pb+Pb collisions at $\sqrt{s}=2.76$~TeV, see figure \ref{fig:RAAv2}. The $K$-factors for each interaction mechanism are tuned such that the experimental $R_{\rm AA}$ is reproduced within the errors above $p_T\simeq10$~GeV. This is possible for all three interaction mechanisms, although in figure \ref{fig:RAA} one is inclined to find a slightly better agreement for the purely collisional energy loss scenario. The $v_2$, figure \ref{fig:v2}, is also reasonably well reproduced by all three scenarios including their respective $K$-factors. Again, the purely collisional interaction is slightly closer to the central data points. One expects, however, a significant contribution to the D-meson $v_2$ from the final hadronic phase, which is not included in the current study.

\section{Azimuthal correlations}

\begin{figure}[tb]
 \centering

   \subfigure{\label{fig:notptcut}\includegraphics[width=0.48\textwidth]{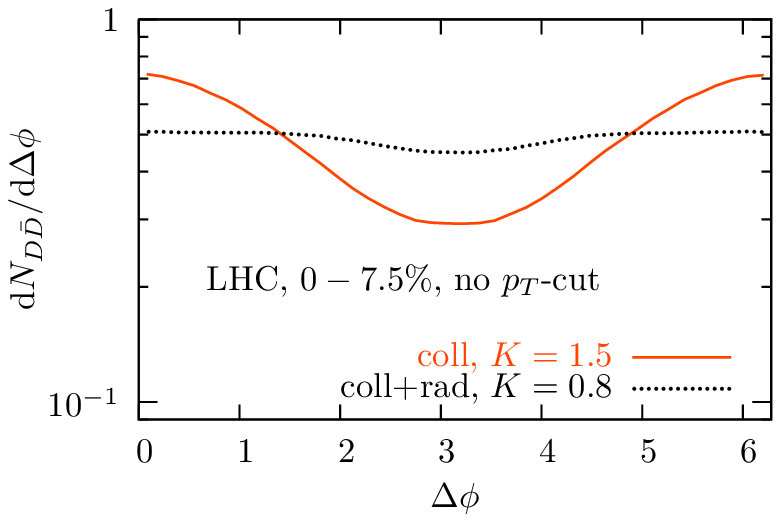}} \hfill
   \subfigure{\label{fig:ptcut}\includegraphics[width=0.48\textwidth]{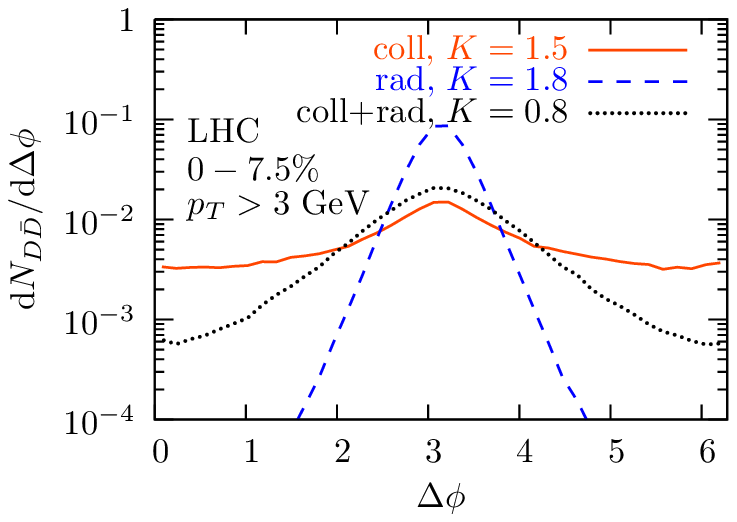}}
 \caption{Azimuthal correlations of $D\bar{D}$-pairs with and without a $p_T$-cut.}
 \label{fig:correl}
\end{figure}

Due to the next-to-leading order (NLO) production processes already in proton-proton collisions the azimuthal distributions will show a broadening of the back-to-back peak at $\Delta\phi=\pi$ and in particular the gluon splitting process leads to an additional enhancement at $\Delta\phi=0$. The theoretical and experimental investigations for $D\bar{D}$ azimuthal correlations even in elementary reactions do not constitute conclusive evidence on the precise shape for initial $c\bar{c}$ azimuthal distributions. The higher mass of the bottom quarks makes similar calculations more easy. In a recent publication \cite{Nahrgang:2013saa} we showed that by using $b\bar{b}$ initial distributions from the NLO+parton shower event generator MC@NLO \cite{mcatnlo} we obtained the same qualitative results as in a simple leading-order back-to-back initialization. In the present study we still initialize the $c\bar{c}$-pairs according to the $p_T$-distributions from FONLL \cite{fonll} and strictly back-to-back. 

In figure \ref{fig:correl} the azimuthal correlations of $D\bar{D}$-pairs, which were initially produced together, are shown for all pairs in the left plot \ref{fig:notptcut} and for pairs, where both transverse momenta are above $3$~GeV in the right plot \ref{fig:ptcut}. One clearly sees that the overall yield is dominated by the low-$p_T$ pairs, which show signs of thermalization with the flowing background medium -- depending on the interaction mechanism. While the distribution is flat for the collisional+radiative scenario, one even observes a final correlation around $\Delta\phi=0$ for the purely collisional scenario, which stems from the radial flow of the medium, pushing a $c\bar{c}$-pair towards smaller opening angles. This is the so-called ``partonic wind'' effect \cite{Zhu:2007ne}. When performing a cut in $p_T$ to concentrate on intermediate and higher $p_T$ this picture is confirmed: the purely collisional interaction is most efficient in washing out the initial correlations during the evolution in the medium. The purely radiative scenario is least efficient, the residual correlation at $\Delta\phi=\pi$ are most pronounced over a small background of isotropized pairs. This is a natural consequence of the larger average $p_\perp^2$ for the purely collisional energy loss mechanism than for any other of the energy loss mechanisms as discussed above.

We thus conclude that a precise measurement of heavy-flavor azimuthal correlations will give insight into the nature of the interaction between heavy-quarks and the QGP beyond what we know from the traditional observables. Implementing more realistic exclusive initial distributions and studying quantitative correlation observables, which are closer to the experimentally accessible ones, is work in progress.

\ack We are grateful for support from  the Hessian LOEWE initiative Helmholtz International 
Center for FAIR, ANR research program ``hadrons @ LHC''  (grant ANR-08-BLAN-0093-02),  TOGETHER project R\'egion Pays de la Loire and I3-HP.

\end{document}